\documentstyle[aps]{revtex}
\begin{document}
\draft
\title{
CONSTITUENT QUARK STRUCTURE AND $F_{2}^{p}$ DATA                  
}

\author{Firooz Arash$^{(a,b)}$\footnote{e-mail: arash@vax.ipm.ac.ir} and
Ali Naghi Khorramian$^{(c)}$}
\address{Abdus Salam International Center for Theoretical Physics, Trieste, Italy}
\address{
$^{(a)}$
Institute for Studies in Theoretical Physics and Mathematics
 P.O.Box 19395-5531, Tehran, Iran \\
$^{(b)}$
Physics Department, Shahid Beheshti University, Tehran, Iran 19834 \\
$^{(c)}$
Physics Department, Amir Kabir University of Technology,
Hafez Avenue, Tehran, Iran \\
}
\date{\today}
\maketitle
\begin{abstract}
We have calculated the partonic structure of a constituent quark in the leading order in QCD for the first time and examined its implications on the proton 
structure function $F_{2}^{p}$ data from HERA. It turned out that although qualitatively it agrees with the data but for a finer refinement we need to consider an {\it{inherent}} component due to gluons which act as the binding agent between constituent quarks in the proton. Good agreement with
nucleon structure function, $F_{2}(x,Q^{2})$, for a wide range of
$x=[10^{-6},1]$ and $Q^{2}=[0.5,5000] GeV^{2}$ is reached.    
{\bf PACS Numbers  13.60 Hb, 12.39.-x, 13.88 +e, 12.20.Fv}
\end{abstract}
\maketitle

Our understanding of hadronic structure is based on two ingredients:(i) The hadronic spectroscopy; which was the original motivation for the introduction of quarks. In this discipline the quarks are massive particles and their bound states describe the static properties of hadrons. Quarks of this domain are usually refered to as Constituent Quarks (CQ). (ii) The Deep Inelastic Scattering (DIS) data is the other source of information about the hadronic structure. The interpretation of DIS data relies upon the quarks of the QCD Lagrangian, current quarks, which have a very small mass. The two kind of quarks are not only different in their masses, but there are other important differences in their other properties; for example, the color charge of quark field in QCD Lagrangian in not gauge invariant, whereas the color associated with the constituent quark is a well defined entity\cite{1}. \\
The constituent quark is defined as a quasi-particle emerging from the dressing of valence quark with gluons and $q-\bar{q}$ pairs in QCD. It is, however, not so easy to pin down the process of dressing itself. For the construction of such an object requires that it must carry color and spin among other quantum numbers and needs to be in conformity with the color confinement. Such an object has been studied recently within the framework of field theory\cite{2} and its emergence from QCD is established. \\
The concept of using CQ as an intermediate step between the quarks of QCD Lagrangian and hadrons is not new. Altarelli and Cabibbo and co-workers used it in the context of a broken $SU(6)$x$O(3)$ long before\cite{3}, R.C. Hwa in his elaborated work,terming a CQ as {\it valon}, extended it and showed its application to many physical processes \cite{4}. In Ref.[1] it was suggested that the concept of dressed quark and gluon might be useful in the area of jet physics and the heavy quark effective theory. Although the concept and the use of CQ has been around for quite sometimes, but no one actually calculated its partonic contents. The goal of this paper is to calculate the structure function of a CQ and try to examine its applicability to the structure function $F_{2}$ of proton, for which there are ample data from DIS experiments covering a wide range of kinematics both in $x$ and $Q^2$. Since, by definition a constituent quark has identical structure in every hadron, once its structure is established, in principle it would permit to calculate the structure of any hadron. In doing so, we will follow the philosophy outlined in Ref.[4]; that is, in a DIS experiment, at high enough $Q^2$, it is the structure of CQ which is being probed and at sufficiently low $Q^2$ this structure cannot be resolved and a CQ behaves as valence quark. In this picture, partons "observed" in DIS experiments are considered as the components of CQ. 
Under the assumption given above, in the high $Q^2$ DIS experiment the structure function of a CQ can be written as:
\begin{equation}
F_{2}^{U}(z,Q^2)=\frac{4}{9}z(G_{\frac{u}{U}}+G_{\frac{\bar{u}}{U}})+ \frac{1}
{9}z(G_{\frac{d}{U}}+G_{\frac{\bar{d}}{U}}+G_{\frac{s}{U}}+G_{\frac{\bar{s}}{U}})+...
\end{equation}
where all the functions on the right-hand side are the probability functions for quarks having momentum fraction $z$ in a $U$-type CQ at $Q^2$. Similar expression can be written for the $D$-type CQ. These functions are calculated using moment space representation. Moments are defined as
\begin{equation}
M_{2}(n,Q^2)=\int_{0}^{1}z^{n-2}F_{2}(z,Q^2) dz
\end{equation}
\begin{equation}
M_{\Omega}(n,Q^2)=\int_{0}^{1}z^{n-1}F_{\Omega}(z,Q^2) dz
\end{equation}
where $\Omega$ stands for valence, singlet and non-singlet partons in a CQ. These moments in the leading order in QCD are calculated and are given as:
\begin{equation}
M^{u,d(val)/CQ}(n,Q^2)=M^{NS}=exp[-d_{NS}t]
\end{equation}
\begin{equation}
M^{sea/CQ}(n,Q^2)=\frac{1}{2f}[M^{S}-M^{NS}]
\end{equation}
where, $f$ stands for number of active flavors, and $t$ is the evolution parameter defined as
\begin{equation}
t={\it{ln}}\frac{\it{ln}\frac{Q^2}{\Lambda^2}}{\it{ln}\frac{Q_{0}^{2}}{\Lambda^2}}
\end{equation}
We have taken $Q^{2}_{0}=0.215$  $Gev^{2}$ and $\Lambda=0.22$ Gev. 
$M^{NS}$ is the non-singlet moment and $M^{S}$ is the singlet one given as:
\begin{equation}
M^{S}=\frac{1}{2}(1+\rho)exp(-d_{+}t)+\frac{1}{2}(1-\rho)exp(-d_{-}t)
\end{equation}
$d_{NS}$, $d_{+(-)}$, and $\rho$ are anamolus dimensions given in Ref [5].
In Figure 1, we have shown these moments for several $Q^2$ values. Once the moments are calculated, the associated distribution functions can be obtained using inverse Mellin transformation techniques, though after going through tedious calculations. For the valence sector the following expression is obtained:
\begin{equation}
zu_{v/CQ}(z,Q^2)=zd_{v/CQ}(z,Q^2)=a z^b (1-z)^c
\end{equation}
where $a$, $b$, and $c$ are functions of $t$.
This distribution function for the valence quark in CQ satisfies the usual number sum rules:$\int_{0}^{1}u_{v}(z,Q^2)dz=1$ for all $Q^2$'s.
For the sea quark distribution in a CQ, the following form is obtained:
\begin{equation}
zq_{sea(CQ)}(z,Q^2)= z^{\beta}(1-z)^{\gamma}(\alpha z^{0.5}+\delta z +\psi)
\end{equation}
Again, $\beta$, $\gamma$,... are functions of $t$. In Figure 2, the parton distribution in a CQ is depicted for several $Q^2$.
Of course, there is no experimental data on the structure function of a CQ to check against the above results. The best we can do is to use convolution theorem and attempt to calculate the proton structure function using eqs.(8,9). To put it differently; we convolute the CQ structure function with the CQ distribution in a proton and sum over all three CQ in the proton:
\begin{equation}
F_{2}^{p}(x,Q^2)=\sum_{CQ}\int_{x}^{1}dy G_{\frac{CQ}{p}}(y)F^{CQ}_{2}(\frac{x}{y},Q^2)
\end{equation}
where $G_{\frac{CQ}{p}}(y)$ is the probability of finding a CQ with momentum fraction $y$ in a proton. They are given in eqs.(14-15), bellow [4]. The summation runs over all constituent quarks of the hadron. In Figures 3 and 4 we present various parton distributions in a proton using the above procedure. As it is evident from Figure 5, going one step forward to the structure function of proton, $F_{2}^{p}$ we notice that these results fall a few percent short of representing  the actual data. We attribute this shortcoming to the fact that in the formation of bound state, a CQ can emit gluon which in turn decays into $q$-$\bar{q}$ pairs, therefore there should be a residual component to the partons in a hadron, we call this component the {\it{inherent}} partons and its contribution can be calculated using splitting functions. Event hough this component is intimately related to the bound state problem, and hence it has a non-perturbative origin, but for not so small values of $Q^2$ we have calculated the process of
$CQ\rightarrow CQ + gluon\rightarrow q-\bar{q}$ perturbatively at an initial 
$Q_{CQ}^2=0.65$ where $\alpha_{s}$ is still small enough. The corresponding splitting functions are as follows:
\begin{equation}
P_{gq}(z)=\frac{4}{3}\frac{1+(1-z)^2}{z}
\end{equation}
\begin{equation}
P_{qg}(z)=\frac{1}{2}(z^2+(1+z)^2)
\end{equation}
The joint probability distribution for the process at hand we get:
\begin{equation}
q_{inh}(x,Q^2)=\bar{q}_{inh}=N\frac{\alpha^{2}_{s}}{(2\pi)^2}\int_{x}^{1}\frac{dy}{y}P_{qg}(\frac{x}{y})\int_{y}^{1}\frac{dz}{z}P_{gq}(\frac{y}{z})G_{CQ}(z)
\end{equation}
$N$ is a factor depending on $Q^2$ and $G_{CQ}$ is the distribution of CQ in proton. We borrow from \cite{4}:
\begin{equation} 
G_{U/p}(y)=7.98y^{0.65}(1-y)^2 
\end{equation}
\begin{equation}
G_{D/p}(y)=6.01y^{0.35}(1-y)^{2.3}
\end{equation}
Now, adding this last contribution to the see quark distributions emerged from the CQ structure, we can see from Figure 5 that it can reproduce experimental data \cite{6} on $F^{p}_{2}$ rather nicely, even in the Leading-Order, for a wide range of $Q^2$ as low as $0.5$ $Gev^{2}$ and all the way up to several thousand $Gev^{2}$, and for $x$ down to $10^{-6}$. For the purpose of comparison we have also shown the Next-to-Leading-Order solution of GRV \cite{7} results. One last point deserves to be addressed here is the incorporation of $\bar{u}\not=\bar{d}$ for the proton. We have not taken this breaking of $SU(2)$ in the nucleon sea in to consideration here; because there is no asymmetry for the creation of light flavor sea within the CQ. The asymmetry is specific for the hadron, for which the structure of a CQ is common. This asymmetry at the hadronic level can be incorporated in this model by requiring that an {\it{inherent}} $q$ or
$\bar{q}$ to recombine with a CQ and make a meson-baryon bound state. That means a nucleon may fluctuate to $\pi$$N$ and $\pi\Delta$ bound states, but this is very similar to the notion of meson cloud models of nucleon and will not be addressed here. \\ 
In conclusion, For the first time we have calculated the internal structure of a constituent quark explicitely in the leading order in QCD and examined its applicability to the real data on proton structure function, $F_{2}^{p}$. We have found that to describe the data one needs to consider an extra component in the sea sector which is attributed to the formation of CQ bound states in the hadron.  \\
{\it{Acknowledgment}}: We thank the Abdus Salam ICTP, for warm haspitality of the center, where this work was carried out in part.   \\\\
{\bf{Appendix}}
In this appendix we give the numerical values for the parameters appearing in 
eqs. (8, 9, 13) of the text at a typical value of $Q^{2}=0.65$ $GeV^{2}$:\\
For the valence quark distribution of eq. 8:\\
$a=0.348$, $b=1.238$, $c=-0.672$. \\
For the sea distribution of eq. 9 we have: \\
$\alpha=-0.0155$, $\beta=-0.073$, $\gamma=1.097$, $\delta=0.0084$, and $\psi=0.0086$; and for the normalization factor, $N$, appearing in eq. 13 we have: $N=0.65$. 

\newpage 
{\bf{Figure Caption}} \\
\\
Figure-1. Moments of partons in a CQ at various $Q^{2}$. \\
Figure-2. Parton distributions in a CQ at several $Q^{2}$.\\
Figure-3. Prediction of the model for valence distribution in proton. \\
Figure-4. Prediction of the model for the  sea quark distribution in proton is depicted. Notice that the thin line is that of the CQ and the dotted line is the contribution of {\it{inherent}} partons. The thick line is the sum of the two. \\
Figure-5. Proton structure function $F_{2}^{p}$ as a function of $x$ calculated using the model ({\it{thick line}}) and compared with the data from
Ref. [6] for different $Q^{2}$ values. The {\it{thin line}} is the prediction of GRV Ref.[7]. \\
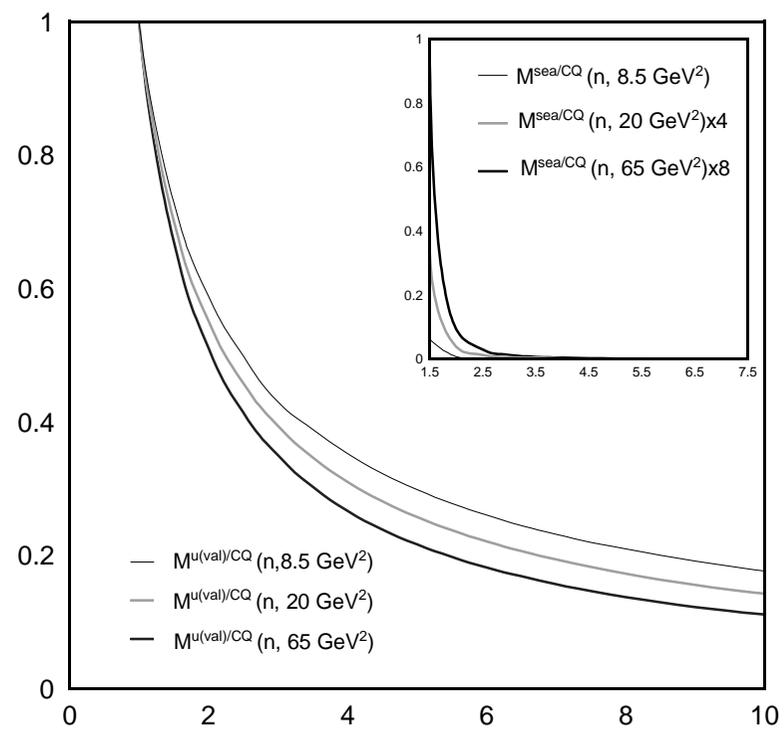
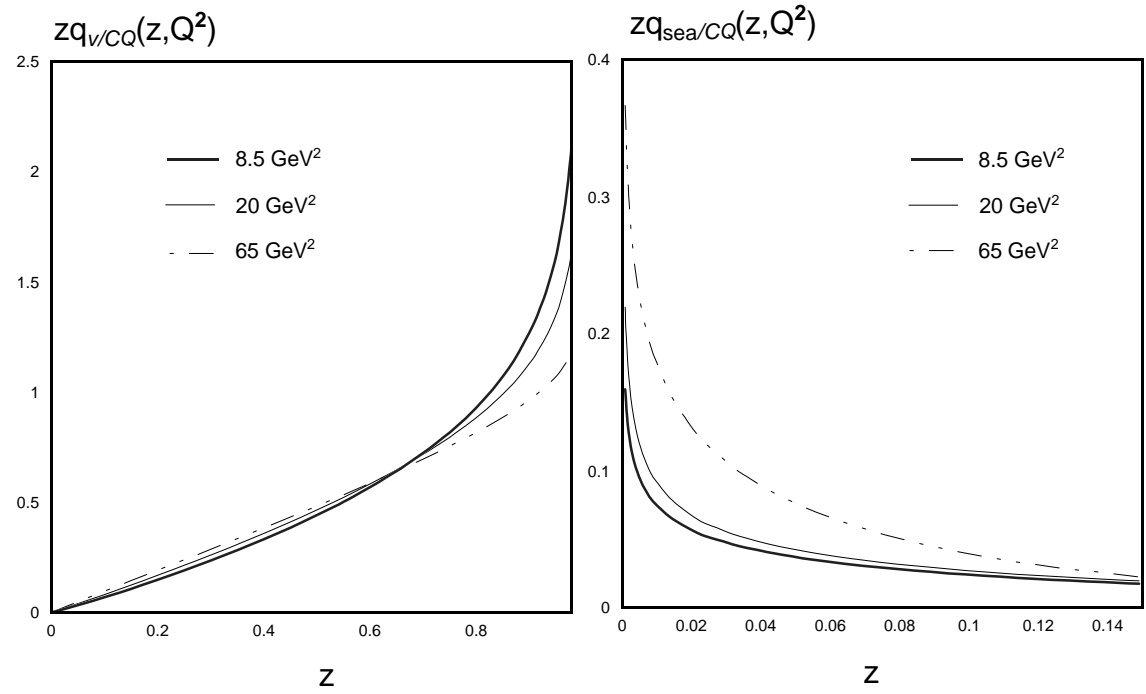
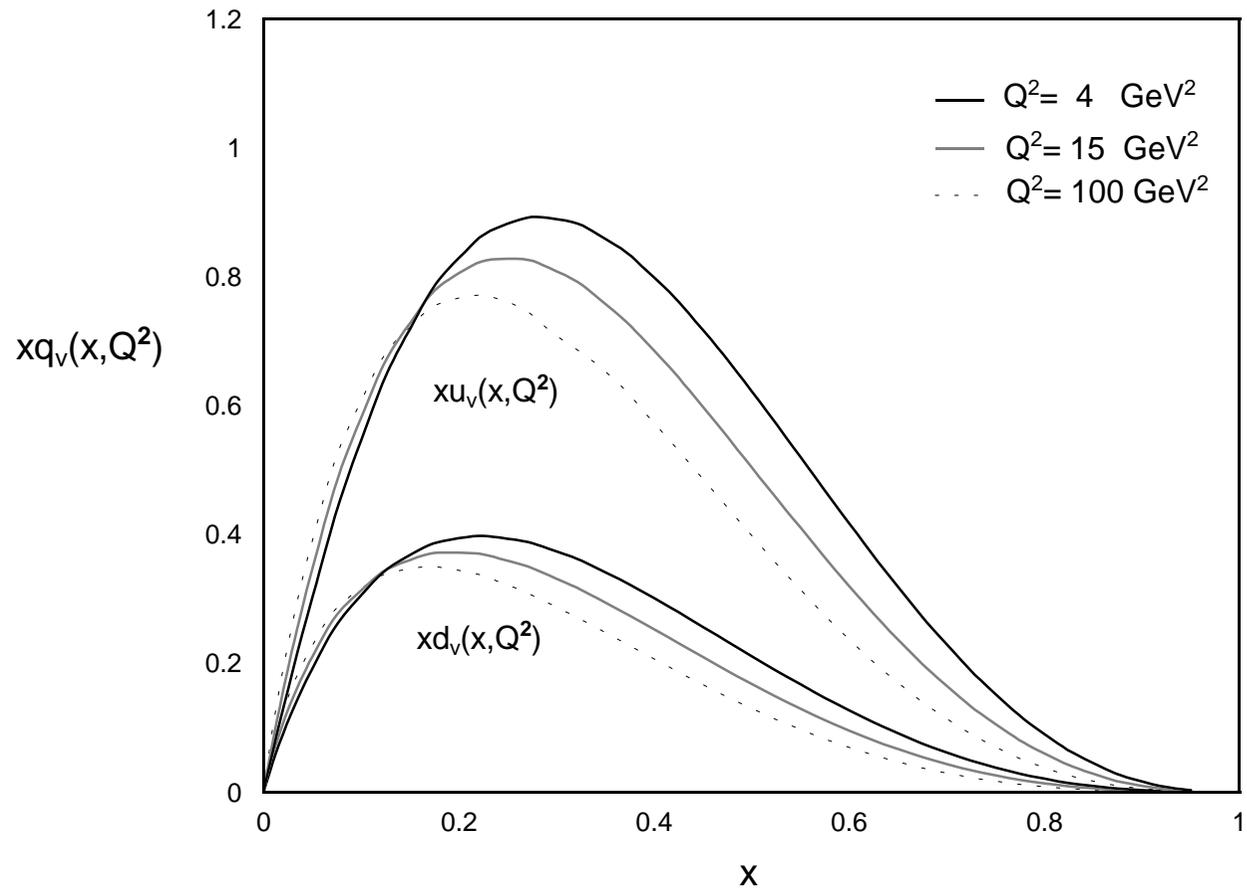
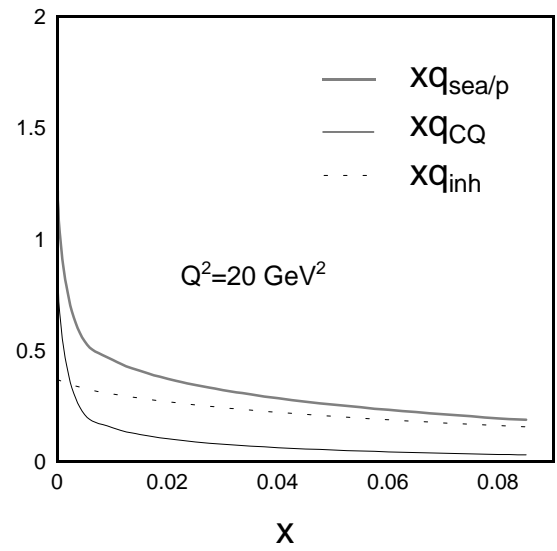
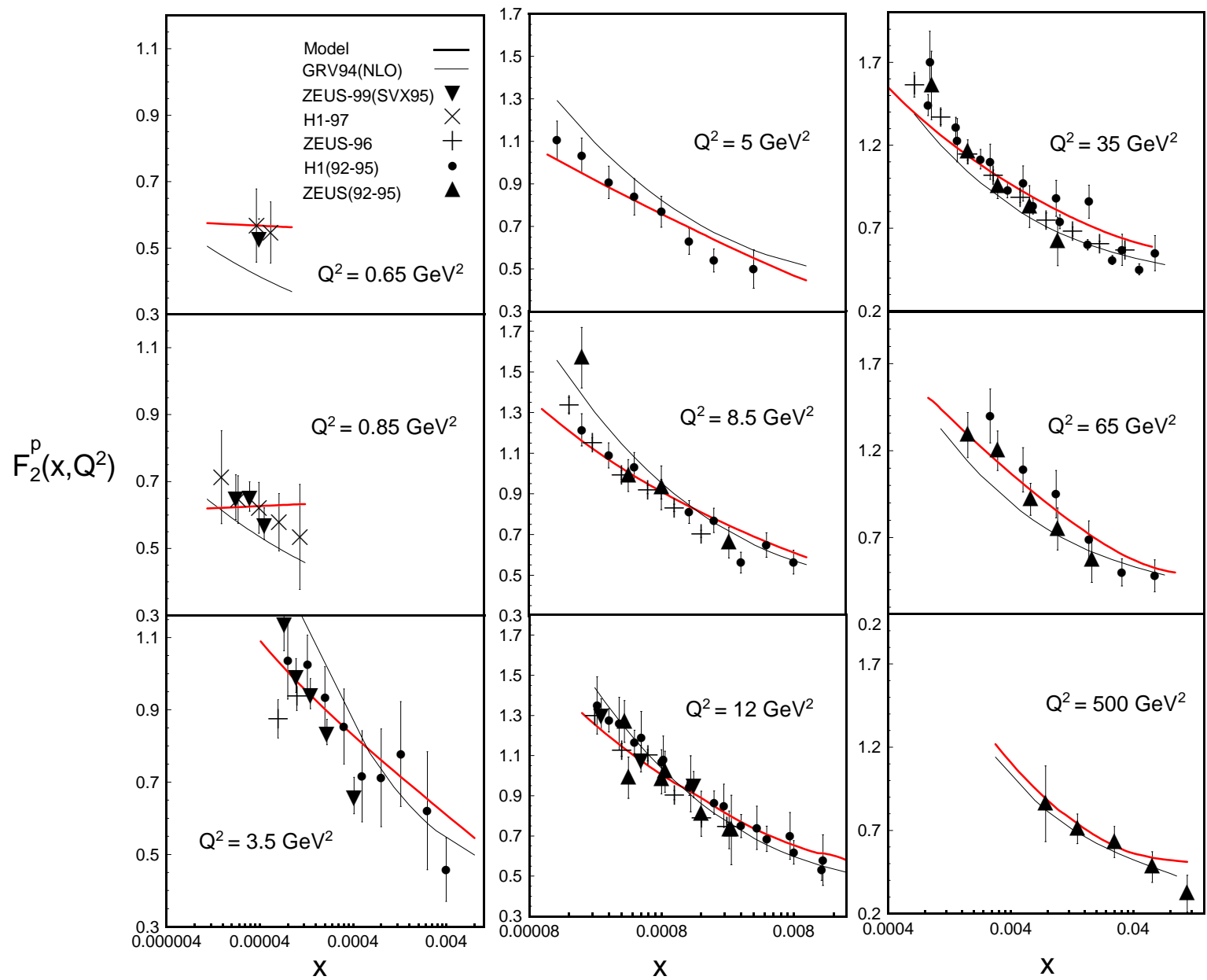
\end{document}